\documentclass[twocolumn,trackchanges]{aastex701}

\usepackage{amsmath}

\shorttitle{BCG and ICL Separation}
\shortauthors{Contini and Yi}

\begin{document}

\title{On the Separation Between the Brightest Cluster Galaxy and the Intracluster Light}

\author{Emanuele Contini}
\altaffiliation{Yonsei University}
\affiliation{Department of Astronomy and Yonsei University Observatory, Yonsei University, 50 Yonsei-ro, Seodaemun-gu, Seoul 03722, Republic of Korea}
\email[show]{emanuele.contini82@gmail.com}

\author{Sukyoung K. Yi}
\altaffiliation{Yonsei University}
\affiliation{Department of Astronomy and Yonsei University Observatory, Yonsei University, 50 Yonsei-ro, Seodaemun-gu, Seoul 03722, Republic of Korea}
\email[show]{yi@yonsei.ac.kr}

\begin{abstract}
We present a model-based framework to define an aperture separation between the brightest cluster galaxy (BCG) and the intracluster light (ICL). Using the intrinsic BCG and ICL components predicted by the semi-analytic model \texttt{FEGA25}, we determine, for each halo, the aperture radius that minimizes the bias in the recovered ICL mass. The optimal radius is obtained by balancing the BCG stellar mass lying outside the aperture with the ICL mass enclosed within it. At $z=0$, the optimal aperture in physical units increases with halo mass approximately as $r_{\rm cut}\propto M_{\rm halo}^{0.28}$, close to the expected virial scaling. When expressed in units of the virial radius, the aperture becomes nearly independent of halo mass, with $r_{\rm cut}/R_{\rm vir}\simeq 0.045$ at $10^{14}M_\odot$. Applying the resulting aperture prescriptions, we recover the intrinsic BCG mass to better than 8\% (lowest percentile), while the ICL mass is recovered with median values very close to unity, and an average scatter of $\pm 3\%$. Extending the analysis to $z=1$ and $z=2$, we find that the mass trends remain similar, with both the slope and intercept of the $r_{\rm cut}/R_{\rm vir}$ relation remaining stable. Robustness tests show that the inferred aperture is most sensitive to the ICL concentration and BCG size, but the main trends are preserved. Our results provide a physically motivated aperture definition for connecting observational BCG--ICL measurements to intrinsic model components.
\end{abstract}



\section{Introduction}
\label{sec:intro}

The intracluster light (ICL) is the diffuse stellar component that occupies the space between galaxies in groups and clusters (\citealt{gonzalez2013,contini2014,mihos2017,iodice2017,tang2018}, among many
others). It is generally associated with stars that are no longer gravitationally bound to any individual galaxy, but instead orbit within the global potential of the host dark matter halo. During the last two decades, several observational and theoretical studies have investigated the origin of the ICL, the mechanisms responsible for its assembly, its stellar properties, and its connection with the dynamical evolution of groups and clusters (see the reviews by \citealt{contini2021,montes2022,arnaboldi2022,contini2024c}, and references therein).

Observationally, the ICL is extremely faint and is usually embedded in the extended light distribution of the brightest group or cluster galaxy (BGG/BCG; hereafter BCG). Its low surface brightness ($<26$ mag arcsec$^{-2}$ in the optical range), often comparable to or below the sky background, makes its detection challenging (e.g. \citealt{montes2014}), especially at high redshift. After sky subtraction, one must remove the contribution of satellite galaxies (e.g. \citealt{presotto2014}) and, most importantly for the present work, separate the diffuse light from the outer envelope of the BCG (e.g. \citealt{burke2015,iodice2017,demaio2018,spavone2018,ragusa2022}).

The physical properties of the ICL provide important information on the assembly history of the host halo. In \cite{contini2023}, we reviewed the main formation mechanisms discussed in the literature and identified three relevant channels: stellar stripping of satellite galaxies (\citealt{rudick2009,rudick2011,martel2012,contini2014,contini2018,demaio2015,demaio2018,montes2018}), mergers between satellites and the BCG
(\citealt{monaco2006,murante2007,contini2014,burke2015,groenewald2017,joo2023}), and the accretion of preprocessed ICL formed in other halos (\citealt{mihos2005,sommer-larsen2006,contini2014,ragusa2023}). Stripping and mergers produce diffuse light directly within the virial radius of the host halo, whereas preprocessing refers to ICL that formed in progenitor systems and was later accreted during hierarchical growth. A detailed discussion of this distinction is given in \cite{contini2023}. In the same work, we also explored the link between ICL formation and halo dynamical state by studying the role of halo concentration, assuming an Navarro-Frenk-White (NFW) density profile \citep{nfw1996}. We found that more concentrated halos tend to host larger ICL fractions, suggesting that halo concentration is a key driver of ICL assembly.

A central difficulty in ICL studies is that the BCG and ICL are spatially embedded. Their surface-brightness profiles often connect smoothly, making any operational definition of the BCG--ICL boundary method dependent. This has led to several definitions in the literature, each probing a different aspect of diffuse light \citep{zibetti2005,burke2015,iodice2017,demaio2018,montes2018,zhang2019,kluge2021,werner2023,brough2024,mayes2026}. A common approach is to apply a surface-brightness threshold and assign all light fainter than a given limit (see above) to the ICL \citep{zibetti2005,burke2015,kluge2021,mayes2026}. This method is simple and closely connected to observations, but the inferred ICL fraction depends on the adopted threshold, photometric band, and data depth. Another possibility is to use fixed physical apertures or radial cuts, assigning the light outside a given radius to the diffuse component \citep{demaio2018,contini2022,brough2024,butler2025}. Such definitions are easy to reproduce, but they do not account for possible variations of BCG size or of the BCG--ICL transition region with halo mass, redshift, and galaxy structure. Aperture-based definitions are also common in simulations: for example, \citet{pillepich2018} used fixed stellar apertures in IllustrisTNG to quantify the stellar mass content of groups and clusters, showing that the inferred contribution of the diffuse component depends on the aperture definition (see also \citealt{contreras-santos2022,contreras-santos2024,brough2024,butler2025,montenegro-taborda2025}).

A different class of methods relies on profile decomposition. In this case, the BCG+ICL surface-brightness distribution is fitted with multiple components, often using de Vaucouleurs or Sersic profiles, and the outer or shallower component is identified with the ICL \citep{gonzalez2005,seigar2007,zhang2019,kluge2021,joo2023}. This provides a continuous description of the light distribution, but the result depends on the
number of components, the adopted functional forms, and the fitted radial range. Other studies define a transition radius, namely the radius where the diffuse component starts to dominate over the BCG contribution
\citep{montes2018,contini2021,chen2022,marini2022}. This definition is physically motivated, although the transition can be broad and system dependent.

Recent observational works have also used non-parametric or image-based techniques, such as multi-galaxy fitting, wavelet decomposition, and low-surface brightness tools designed to separate diffuse emission from galaxies and background structures \citep{presotto2014,jimenez-teja2018,ellien2021,martinez-lombilla2023,brough2024}. These approaches are especially useful when satellite masking, background subtraction, Point Spread Function (PSF) effects, and image depth strongly affect the measurement. In simulations, the availability of phase-space and three-dimensional particle information enables additional definitions, including kinematic, binding-energy, structure-finder, or density-based classifications \citep{dolag2010,puchwein2010,rudick2011,remus2017,brown2024,butler2025,jeon2026}. For example, \citet{brown2024} identify the ICL in Horizon-AGN using the output of \texttt{AdaptaHOP}: star particles within the cluster that are not assigned to any stellar structure are classified as ICL. A related point was recently emphasized by \citet{brown2026}, who compared ICL assembly across different hydrodynamical simulations using a homogenized identification framework. These methods are powerful in simulations, but they are not directly equivalent to photometric definitions based on surface brightness, apertures, or profile decomposition.

Because of these ambiguities, several observational studies choose not to separate the BCG and ICL and instead analyse them as a single BCG+ICL component, especially at high redshift where the ICL is extremely faint
\citep{gonzalez2021,werner2023,golden-marx2025} and the spatial resolution of the data might be poor. This avoids imposing an uncertain boundary, but prevents one from studying the two components separately. Distinguishing the BCG from the intracluster light is important because the two stellar components trace different aspects of cluster assembly. While the BCG primarily reflects the growth of the central galaxy, the ICL is built through the redistribution of stars during mergers, tidal stripping, satellite disruption, and pre-processing, and therefore carries information on the dynamical history of the host halo. This distinction is not merely semantic, since the adopted BCG--ICL boundary directly affects the inferred BCG mass, ICL fraction, their scaling relations with halo mass, and their apparent redshift evolution. At the same time, the two components overlap spatially, making their observational separation intrinsically ambiguous. A physically motivated and reproducible separation scheme is therefore required to connect observational measurements with the intrinsic BCG--ICL decomposition predicted by galaxy-formation models.

In this work, we follow a complementary approach. Since our model provides intrinsic BCG and ICL masses, we use them to define an aperture-based separation and determine the radius at which the contamination of the extracted ICL by BCG stars balances the amount of ICL missed inside the aperture. This connects an observationally motivated aperture definition to the intrinsic BCG--ICL decomposition predicted by the model.

The paper is organized as follows. In Section~\ref{sec:methods}, we describe the semi-analytic model (SAM) used in this work and introduce the aperture-based framework adopted to model the BCG--ICL system and define the optimal separation radius. In Section~\ref{sec:results}, we present our results. We first focus on $z=0$, characterizing the optimal aperture, its dependence on halo mass, and the recovery of the intrinsic BCG and ICL components. We then extend the analysis to $z=1$ and $z=2$ and test the robustness of the results against different structural assumptions. Finally, in Section~\ref{sec:conclusion}, we summarize our main findings and discuss their implications for observational definitions of the BCG--ICL separation. Stellar masses are derived assuming a \citet{chabrier2003} initial mass function, and unless otherwise stated, all masses are $h (=0.68)$-corrected.

\section{Methods}
\label{sec:methods}
In this section we first summarize the main features of the SAM used in this work, \texttt{FEGA25}, focusing on the quantities that are relevant for the present analysis. We then describe how we use the model outputs to construct an idealized description of the BCG--ICL system. In particular, we explain how the stellar mass associated with the central galaxy and with the diffuse component is distributed through analytic profiles, and how an aperture-based separation is used to define the extracted BCG and ICL masses.

\subsection{\texttt{FEGA25} and Catalogs}
\texttt{FEGA25} is an updated version of the semi-analytic framework presented in \cite{contini2024d}. The model includes a revised treatment of both star formation and AGN feedback. In particular, with respect to the previous implementation, \texttt{FEGA25} adopts an extended Kennicutt-Schmidt relation \citep{shi2011} and a more self-consistent connection between black hole growth and AGN activity (see \citealt{contini2025AGN} and \citealt{contini2025}).

The diffuse stellar component in \texttt{FEGA25} is modelled as a physically motivated ICL component that builds up during the hierarchical assembly of dark matter halos. In the model, the ICL is not assigned by construction as a fixed fraction of the stellar mass, but is produced through the cumulative action of several channels associated with the dynamical evolution of satellite galaxies and their host halos.

The first channel is stellar stripping. As satellite galaxies orbit within the potential well of a more massive halo, part of their stellar mass can be removed by tidal forces and transferred to the diffuse component associated with the central galaxy. In some cases, the tidal interaction is strong enough to fully disrupt the satellite, in which case its remaining stellar mass is also added to the ICL. This channel naturally links the growth of the ICL to the orbital evolution and survival time of satellite galaxies.

A second contribution comes from galaxy mergers. During both minor and major mergers between a central and a satellite galaxy, a fraction of the satellite stellar mass is assigned to the ICL component of the remnant central galaxy. In \texttt{FEGA25}, this fraction is drawn around an average value of $20\%$, with a scatter of $\pm 5\%$ \citep{contini2014,contini2026a}. This prescription accounts for the fact that not all stars from the merging satellite are incorporated into the central galaxy; a fraction is instead dispersed into the diffuse light (see, e.g., \citealt{contini2014,contini2023}).

The model also tracks preprocessed ICL. This component corresponds to diffuse stellar mass that was produced in progenitor halos before they were accreted by the final host. Preprocessed ICL is therefore not an independent formation mechanism, but rather records ICL that originated from stripping, mergers, or satellite disruption in smaller systems and was later incorporated into the main halo through hierarchical growth.

Previous applications of the model have shown that stellar stripping is typically the dominant channel for ICL formation, while mergers provide a secondary contribution \citep{contini2018,contini2024a}. However, the relative importance of these channels can depend on the adopted definition of mergers and on the way in which disrupted satellites are treated \citep{contini2018}. In the present work, we use the ICL masses predicted by \texttt{FEGA25} as the intrinsic diffuse stellar component of each halo. These masses, together with the corresponding BCG stellar masses and halo properties, provide the starting point for the aperture-based modelling described below.

Throughout this work, we include the stellar halo in the diffuse stellar component and use the term ICL to refer to this total diffuse mass. We stress that this is an operational choice and does not imply that the stellar halo and the ICL are physically identical components. Rather, their spatial overlap and related accretion-driven origins make their separation intrinsically definition dependent. Since our aim is to separate the central BCG from the surrounding diffuse stellar component, we treat both consistently as part of the diffuse component in the analysis. We refer the reader to \citet{contini2024e} and \citet{contini2026a}  for the physical motivation and implementation of the stellar-halo modelling.

To ensure adequate numerical resolution, we construct our sample by applying a halo-mass selection criterion. At all redshifts considered, halos are required to satisfy a simulation-dependent threshold in $\log M_{\rm halo}$, with masses expressed in $M_{\odot}/h$. The adopted thresholds are
\[
\log M_{\rm halo,min} =
\left\{
\begin{array}{ll}
13.0 & \mathrm{YS200} \\
14.0 & \mathrm{YS300}
\end{array}
\right.
\]
\noindent\hspace{-0.5em} where YS200 and YS300 are the simulations used in this work. They are dark-matter-only cosmological simulations with box sizes of $200$ and $300\,{\rm Mpc}\,h^{-1}$, respectively. Their dark matter particle masses are $3.26\times10^8$ and $2.2\times10^9\,M_\odot h^{-1}$, respectively. For the details about them, we refer the reader to \citet{contini2025AGN}. The same halo-mass thresholds are applied at all redshifts considered. This conservative selection is intended to minimize numerical effects associated with poorly resolved halos.

\subsection{BCG-ICL model}
\label{sec:BCGICLmodel}

\begin{figure}[t!]
\centering
\includegraphics[width=0.45\textwidth]{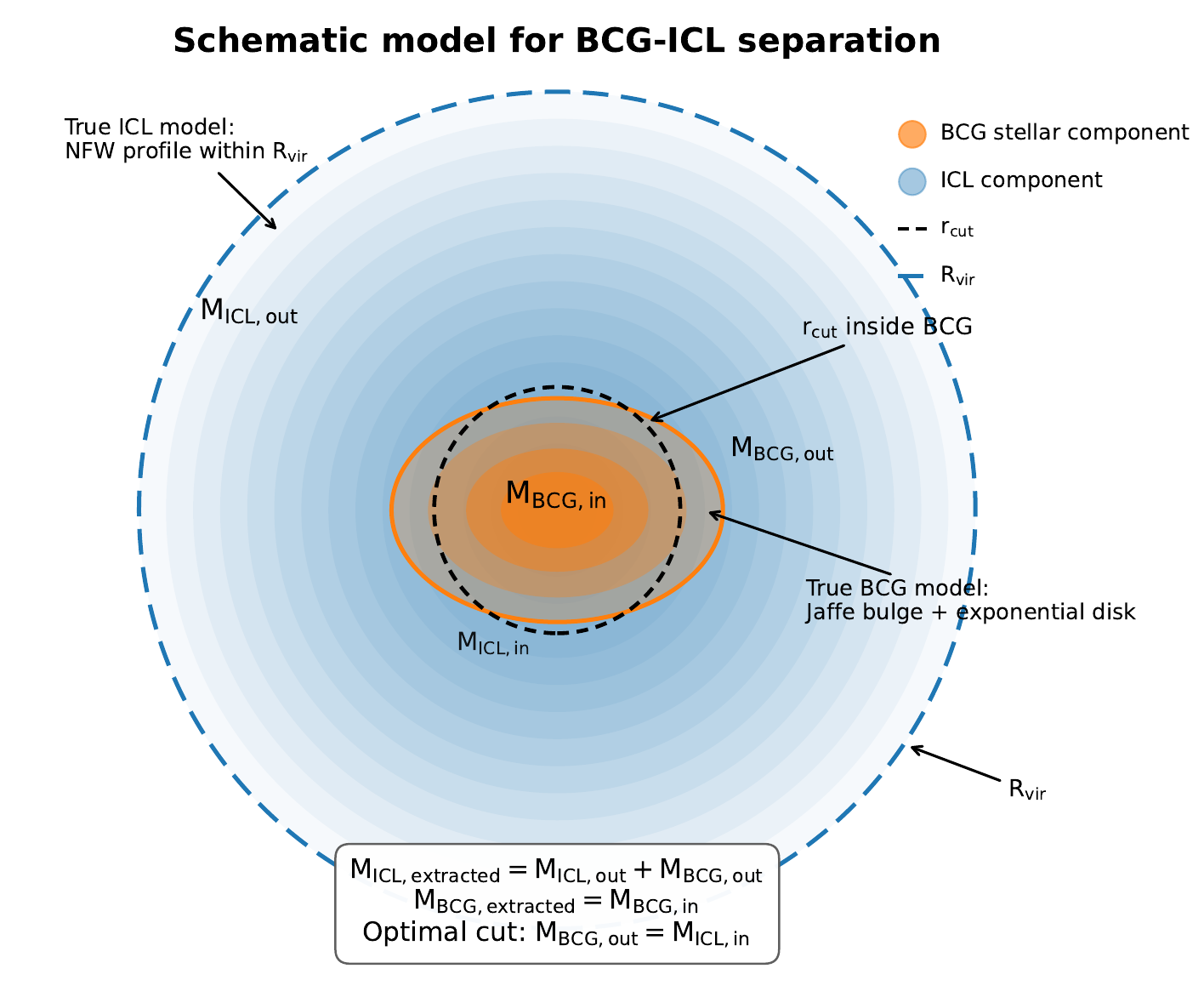}
\caption{Schematic representation of the aperture-based BCG--ICL separation adopted in this work.
The system is modelled as the superposition of a central BCG component and a diffuse ICL component extending within the virial radius, $R_{\rm vir}$. The aperture radius, $r_{\rm cut}$, is allowed to lie within the outer stellar envelope of the BCG. As a result, the extracted ICL contains the true ICL mass outside the aperture plus the fraction of BCG mass lying outside $r_{\rm cut}$, while the extracted BCG corresponds to the true BCG mass enclosed within the aperture. The optimal aperture is defined as the radius where the BCG mass outside the aperture balances the ICL mass inside the aperture. The elliptical appearance of the BCG in the schematic is purely illustrative; all calculations assume spherical, concentric profiles.}
\label{fig:fig0}
\end{figure}

In this section we describe the procedure adopted to define an aperture-based separation between the BCG and the ICL. The basic idea is illustrated schematically in Figure~\ref{fig:fig0}. We model each system as the superposition of two stellar components: a central BCG component and an extended ICL component. The BCG is described as the sum of a Jaffe bulge \citep{jaffe1983} and an exponential disk \citep{freeman1970}, while the ICL is described by an NFW-like profile \citep{nfw1996} normalized within the virial radius, $R_{\rm vir}$.

The separation between the two components is performed through a spherical aperture of radius $r_{\rm cut}$. Material inside this aperture is associated with the central galaxy, while material outside the aperture is classified as diffuse light. Importantly, in our model $r_{\rm cut}$ is allowed to lie within the outer stellar envelope of the BCG. Therefore, a fraction of the true BCG mass may remain outside the aperture and contaminate the extracted ICL component.

For a given aperture, we define
\begin{equation}
M_{\rm ICL}^{\rm ext}
=
M_{\rm ICL}^{\rm out}
+
M_{\rm BCG}^{\rm out},
\label{eq:iclextracted}
\end{equation}
\noindent\hspace{-0.3em}where $M_{\rm ICL}^{\rm out}$ is the true ICL mass outside $r_{\rm cut}$, and $M_{\rm BCG}^{\rm out}$ is the true BCG mass outside $r_{\rm cut}$. The latter term represents the BCG contribution that would be classified as ICL by a purely aperture-based separation.

Conversely, when quantifying the recovery of the BCG component, we define
\begin{equation}
M_{\rm BCG}^{\rm ext}
=
M_{\rm BCG}^{\rm in},
\label{eq:bcgextracted}
\end{equation}
\noindent\hspace{-0.3em}where $M_{\rm BCG}^{\rm in}$ is the true BCG mass enclosed within $r_{\rm cut}$. This definition measures how much of the intrinsic BCG component is recovered by the aperture. We stress that the present approach is based on the intrinsic three-dimensional model components and does not include observational effects such as projection, foreground or background contamination, PSF convolution, sky subtraction, or surface-brightness limits.

The optimal aperture is chosen by requiring the extracted ICL mass to be as close as possible to the true ICL mass. For each halo, we minimize the absolute fractional bias
\begin{equation}
\epsilon_{\rm ICL}
=
\left|
M_{\rm ICL}^{\rm ext}/M_{\rm ICL}^{\rm true}
-
1
\right| .
\label{eq:iclbias}
\end{equation}
Using Equation~(\ref{eq:iclbias}), the optimal aperture can be understood as the radius where two opposite effects balance each other. If the aperture is too small, too much of the BCG lies outside $r_{\rm cut}$, and the extracted ICL is overestimated. If the aperture is too large, too much true ICL lies inside $r_{\rm cut}$, and the extracted ICL is underestimated. The optimal cut for an individual halo is therefore identified by the condition
\begin{equation}
M_{\rm BCG}^{\rm out}
=
M_{\rm ICL}^{\rm in}.
\label{eq:balance}
\end{equation}

We determine the optimal aperture independently for each halo, obtaining one value of $r_{\rm cut}$ per object. We then study how these optimal apertures vary as a function of halo mass and redshift. Since the virial radius scales approximately as $M_{\rm halo}^{1/3}$ at fixed redshift \citep{barkana2001}, we describe the optimal aperture in two complementary ways.

First, we fit the optimal aperture in physical units. For each redshift, we use the simple linear relation
\begin{equation}
y = A + B x .
\label{eq:fitkpc}
\end{equation}
\noindent\hspace{-0.3em}In this case, $y$ is the logarithm of the optimal aperture in kpc, and $x$ is the logarithm of the halo mass normalized to $10^{14} M_{\odot}$. The intercept therefore gives the value of the physical aperture at the pivot halo mass, while the slope describes how the optimal aperture changes with halo mass.

Second, we fit the same optimal apertures in units of the virial radius. In this case we use again a simple linear relation given by Equation~\ref{eq:fitkpc}, but $y$ is the ratio between the optimal aperture and the virial radius, and $x$ is again the logarithm of the halo mass normalized to $10^{14} M_{\odot}$. This second fit tests whether the BCG--ICL separation is better described by a fixed physical aperture or by an aperture that scales with the size of the host halo.

The stellar mass distribution is described using three analytic profiles. The BCG is modelled as the sum of a spheroidal bulge and a stellar disk, while the ICL is modelled as a separate diffuse component.

The bulge component of the BCG follows a Jaffe profile,
\begin{equation}
\rho_b(r) \propto 1/[r^2 (r+a)^2] .
\end{equation}
\noindent\hspace{-0.35em}Here $a$ is the scale radius of the bulge. This profile describes the centrally concentrated spheroidal component of the BCG.

The disk component of the BCG follows an exponential profile,
\begin{equation}
\Sigma_{\rm d}(R) \propto \exp(-R/R_{\rm d}) .
\end{equation}
\noindent\hspace{-0.35em}Here $R_{\rm d}$ is the disk scale length. The total BCG profile is therefore obtained by combining the Jaffe bulge and the exponential disk. Throughout this work, the term ``BCG size'' refers to the radial extent
of the BCG stellar-mass profile, rather than to its stellar mass. In the fiducial bulge--disk model, this spatial extent is set by the Jaffe bulge scale radius $a$ and the exponential-disk scale length $R_d$. Accordingly, the BCG-size variations explored in the robustness tests refer to rescalings of these characteristic radial scales.

The ICL is described by an NFW-like profile,
\begin{equation}
\rho_{\rm ICL}(r) \propto 1/[(r/r_s)(1+r/r_s)^2] .
\end{equation}
\noindent\hspace{-0.35em}The scale radius is related to the virial radius through the ICL concentration, $c_{\rm ICL}=\gamma R_{\rm vir}/\rm r_s$ \footnote{$\gamma$ is a parameter of the SAM and can vary between 1 and 3. See \citet{contini2026a} and references therein for further details on its calibration.}. The profile is normalized so that the total ICL mass inside $R_{\rm vir}$ is equal to the ICL mass assigned to the halo.

For any value of $r_{\rm cut}$, the analytic profiles described above allow us to compute the fractions of BCG and ICL mass lying inside and outside the aperture. The aperture does not modify the intrinsic BCG and ICL components; it only defines how they would be classified by an aperture-based separation. In particular, BCG mass outside $r_{\rm cut}$ contributes to the extracted ICL, whereas ICL mass inside $r_{\rm cut}$ is missed by the extracted ICL.

To estimate the statistical uncertainty of the fitted aperture relations, we use bootstrap resampling. For each bootstrap realization, we resample the halo catalogue with replacement, keeping the same number of objects as in the original sample, and repeat the fit to the $r_{\rm cut}/R_{\rm vir}$ relation. We perform 2000 bootstrap realizations and use the resulting distribution of the fitted parameters to estimate their statistical uncertainty. In the robustness tests, the shaded bootstrap bands correspond to the 16th and 84th percentiles of the resulting distribution. More generally, throughout this work the 16th--84th percentile interval denotes the central 68\% of the relevant distribution and is approximately equivalent to a $\pm1\sigma$ interval for a Gaussian distribution.

In the following section, we use these individual optimal apertures to study how $r_{\rm cut}$ depends on halo mass and redshift. We then derive simple aperture prescriptions, both in physical units and in units of $R_{\rm vir}$, and test how well they recover the intrinsic BCG and ICL masses.

\section{Results and Discussion}
\label{sec:results}\

\begin{figure*}[t!]
\centering
\includegraphics[width=0.94\textwidth]{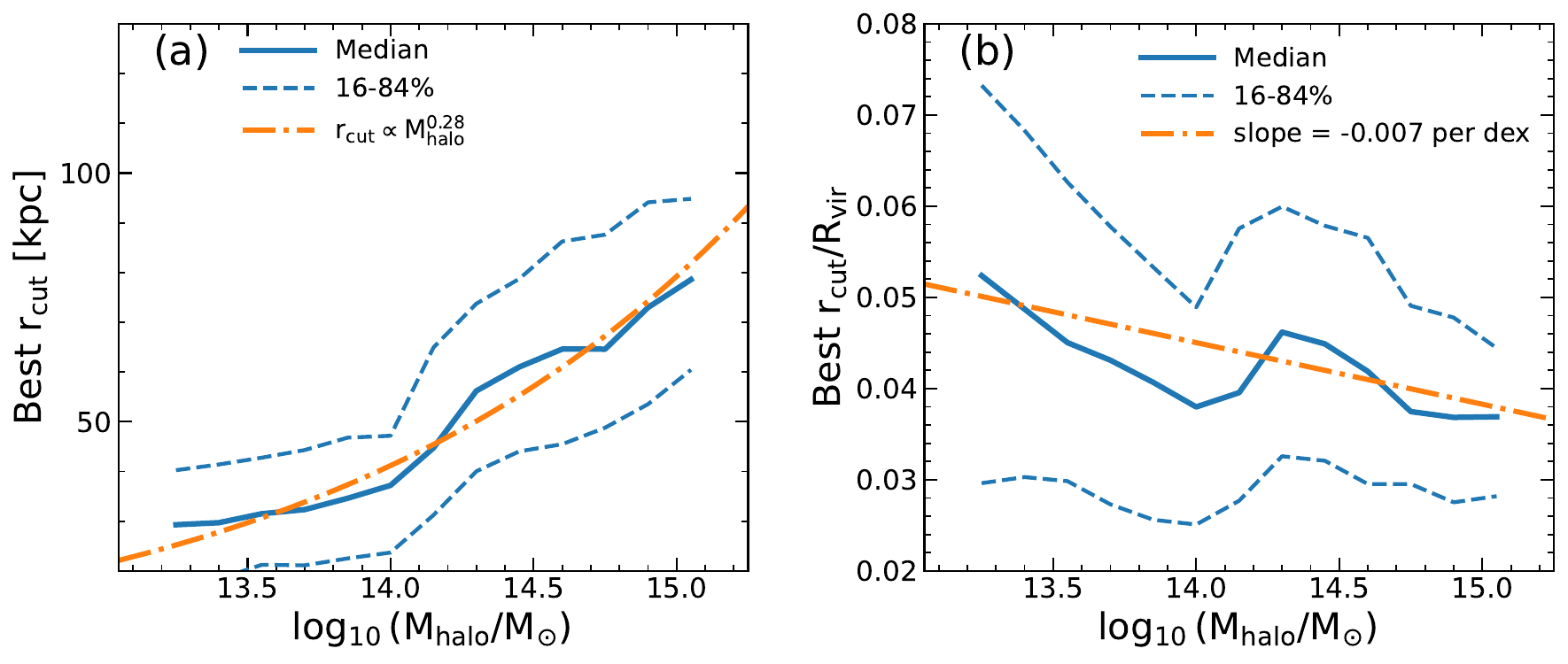}
\caption{Optimal aperture radius as a function of halo mass at $z=0$. Panel (a) shows the best value of $r_{\rm cut}$ in physical units, while panel (b) shows the corresponding aperture normalized to the virial radius, $r_{\rm cut}/R_{\rm vir}$. Solid blue lines show the median relation, and dashed blue lines indicate the 16--84 percentile range. In the left panel, the orange dot-dashed line shows a power-law fit of the form $r_{\rm cut} \propto M_{\rm halo}^{0.28}$, while in the right panel it shows the best-fitting linear trend of $r_{\rm cut}/R_{\rm vir}$ with halo mass, with a slope of $-0.007$ per dex. The weak mass dependence of $r_{\rm cut}/R_{\rm vir}$ indicates that the optimal aperture is approximately proportional to the virial radius, although with a mild decrease toward higher halo masses.}
\label{fig:fig1}
\end{figure*}

\begin{figure*}[t!]
\centering
\includegraphics[width=0.95\textwidth]{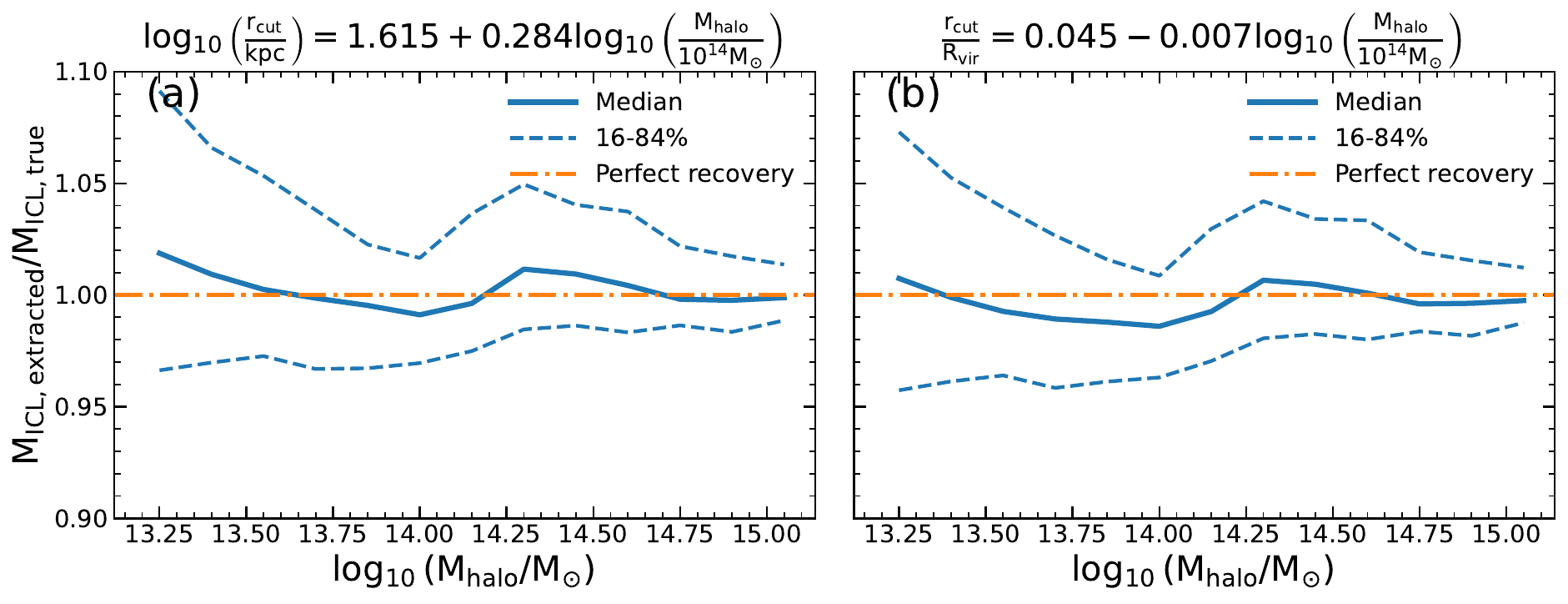}
\caption{Recovery of the ICL mass at $z=0$ obtained by applying the two aperture prescriptions derived from the individual optimal cuts. In each panel, the equation shown at the top specifies how $r_{\rm cut}$ is assigned to each halo as a function of halo mass before measuring the recovered ICL mass. In panel (a), $r_{\rm cut}$ is computed from the best-fitting relation in physical units,
$\log_{10}(r_{\rm cut}/{\rm kpc}) = 1.615 + 0.284 \log_{10}(M_{\rm halo}/10^{14}M_{\odot})$. In panel (b), the aperture is instead computed from the fit in units of the virial radius, $r_{\rm cut}/R_{\rm vir} = 0.045 - 0.007 \log_{10}(M_{\rm halo}/10^{14}M_{\odot})$. Once the aperture is assigned, the extracted ICL mass is measured as the intrinsic ICL mass outside $r_{\rm cut}$
plus the BCG mass lying outside the same aperture. The figure shows the ratio $M_{\rm ICL}^{\rm ext}/M_{\rm ICL}^{\rm true}$ as a function of halo mass. Solid blue lines show the median relation, dashed blue lines indicate the 16--84 percentile range, and the orange dot-dashed line marks perfect recovery.}
\label{fig:fig2}
\end{figure*}

\begin{figure*}[t!]
\centering
\includegraphics[width=0.94\textwidth]{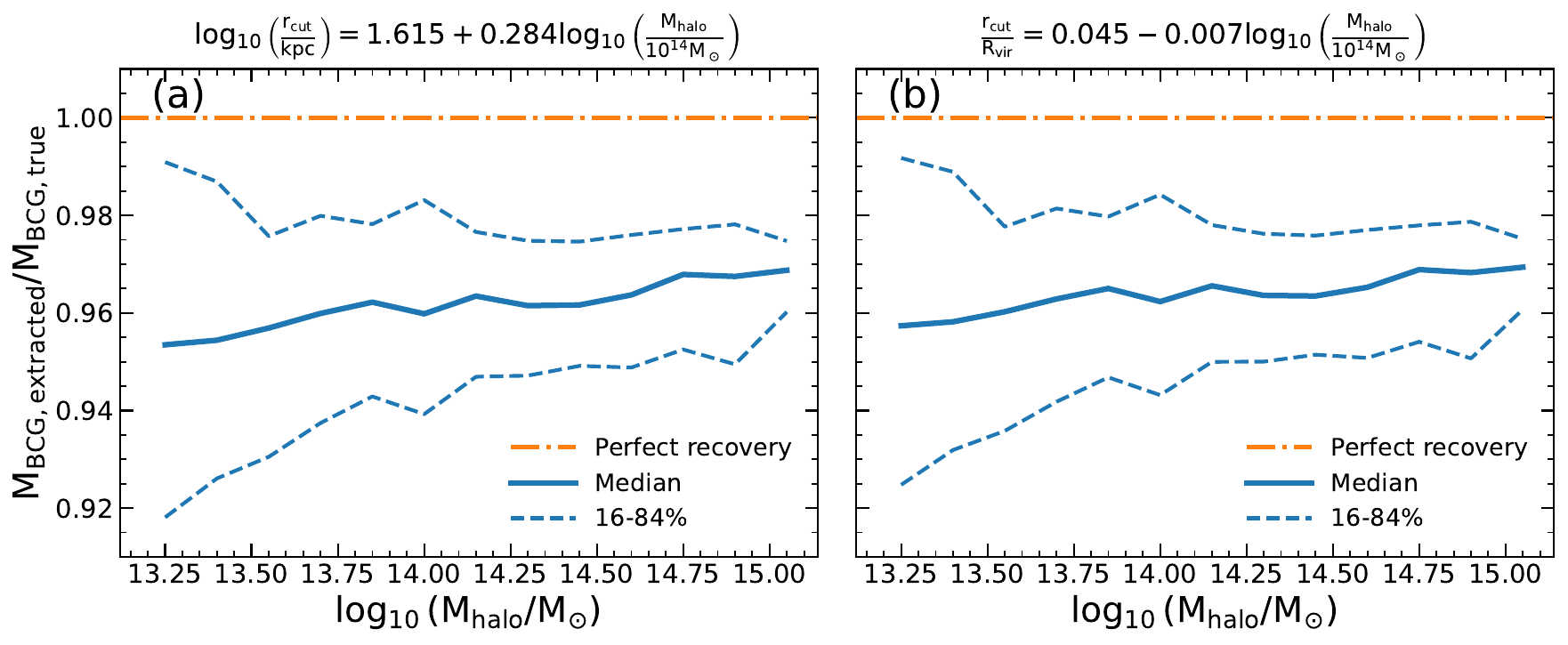}
\caption{Recovery of the BCG mass at $z=0$ obtained by applying the two aperture prescriptions derived from the individual optimal cuts. In each panel, the equation shown at the top specifies how $r_{\rm cut}$ is assigned to each halo as a function of halo mass before measuring the recovered BCG mass. In panel (a), $r_{\rm cut}$ is computed from the best-fitting relation in physical units,
$\log_{10}(r_{\rm cut}/{\rm kpc}) = 1.615 + 0.284 \log_{10}(M_{\rm halo}/10^{14}M_{\odot})$. In panel (b), the aperture is instead computed from the fit in units of the virial radius, $r_{\rm cut}/R_{\rm vir} = 0.045 - 0.007 \log_{10}(M_{\rm halo}/10^{14}M_{\odot})$. Once the aperture is assigned, the extracted BCG mass is measured as the intrinsic BCG mass enclosed within $r_{\rm cut}$. The figure shows the ratio $M_{\rm BCG}^{\rm ext}/M_{\rm BCG}^{\rm true}$ as a function of halo mass. Solid blue lines show the median relation, dashed blue lines indicate the 16--84
percentile range, and the orange dot-dashed line marks perfect recovery.}
\label{fig:fig3}
\end{figure*}

\begin{figure*}[t!]
\centering
\includegraphics[width=0.94\textwidth]{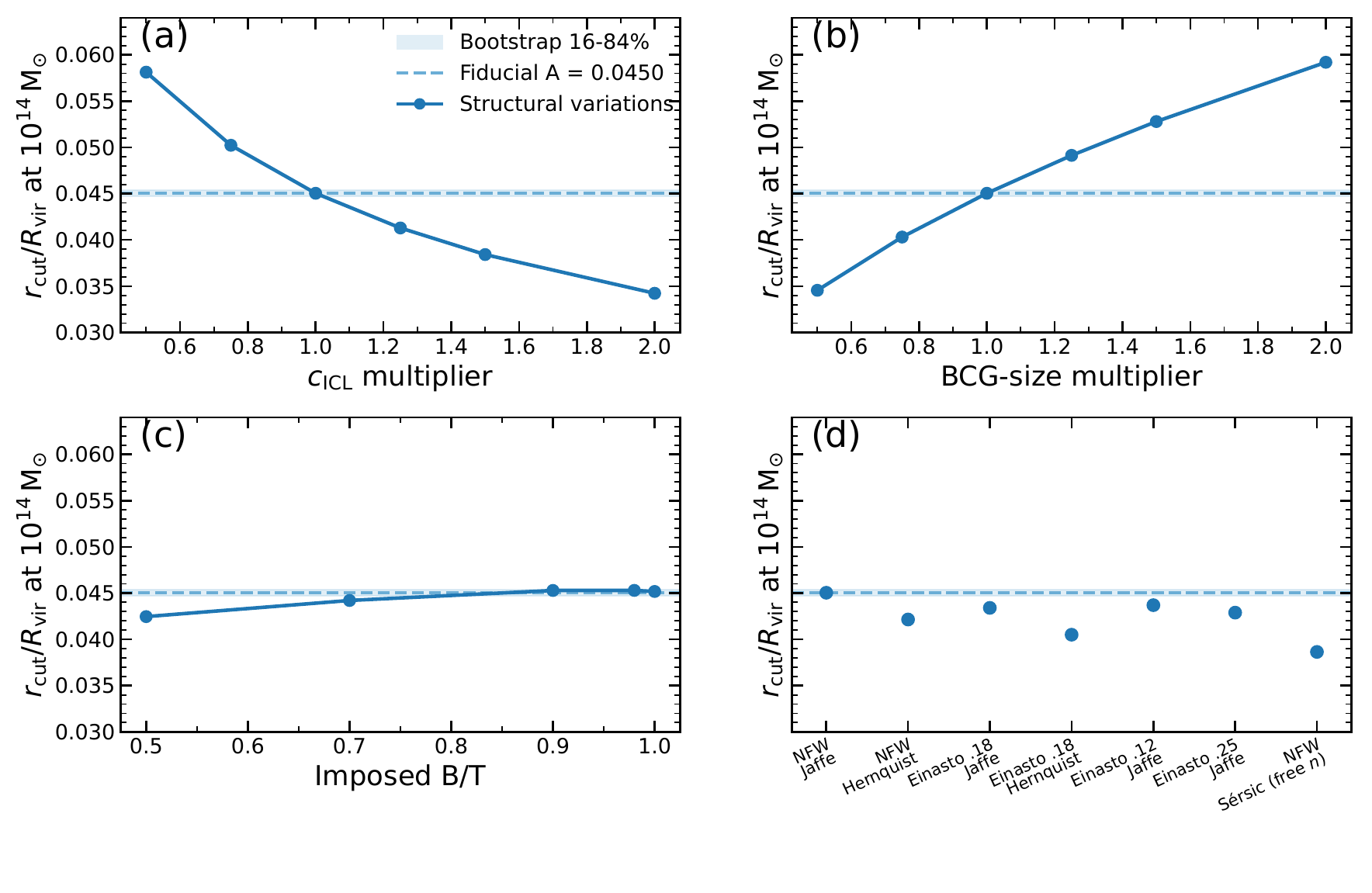}
\caption{Robustness of the optimal aperture with respect to the assumed structural properties of the BCG and ICL components at $z=0$. The quantity shown is the normalization of the scaled-aperture relation, $r_{\rm cut}/R_{\rm vir}$, evaluated at the pivot halo mass $10^{14}M_{\odot}$. The horizontal dashed line marks the fiducial value, $A=0.045$, and the shaded band shows the 16--84 percentile range obtained from bootstrap resampling of the fiducial model. Panel (a) shows the effect of varying the ICL concentration, $c_{\rm ICL}$, relative to the fiducial value. Increasing $c_{\rm ICL}$ makes the diffuse component more centrally concentrated and therefore shifts the optimal aperture to smaller values. Panel (b) shows the effect of rescaling the characteristic size of the BCG. More extended BCG profiles require a larger aperture in order to avoid assigning too much BCG mass to the ICL. Panel (c) shows the dependence on the imposed bulge-to-total ratio, $B/T$, which changes the relative contribution of the spheroidal and disk components to the BCG profile. Panel (d) compares different combinations of BCG and ICL profile shapes, including a single S\'ersic model for the total BCG. Overall, the figure quantifies how sensitive the inferred optimal aperture is to the adopted structural modelling of the BCG--ICL system.}
\label{fig:fig4}
\end{figure*}

We now apply the procedure described in the previous section to determine the optimal BCG--ICL separation aperture and to assess its dependence on halo mass and redshift. We present the results in two steps. First, we focus on the local Universe, at $z=0$, where the available statistics are most robust and the behaviour of the method can be studied in detail. We use this redshift as a reference case to
derive the individual optimal apertures, to fit their dependence on halo mass, and to quantify how well the resulting aperture prescriptions recover the intrinsic BCG and ICL masses. We also use the $z=0$ sample to test the sensitivity of the optimal aperture to the assumed structural properties of the BCG and ICL profiles.

We then extend the same analysis to higher redshift, focusing on $z=1$ and $z=2$. This allows us to investigate whether the aperture prescriptions calibrated at each epoch retain the same qualitative behaviour and to quantify the evolution of the optimal BCG--ICL separation with cosmic time. In particular, we compare physical apertures, expressed in kpc, with halo-scaled apertures, expressed in units of $R_{\rm vir}$. This comparison is useful because the physical size of halos evolves strongly with redshift, while the ratio $r_{\rm cut}/R_{\rm vir}$ directly probes whether the separation between the BCG and the diffuse stellar component follows the growth of the host halo.

\subsection{ICL Recovery in the Local Universe}

Figure~\ref{fig:fig1} shows the dependence of the optimal aperture on halo mass at $z=0$, obtained following the procedure described in Section~\ref{sec:BCGICLmodel}. Panel (a) shows the optimal cut in physical units,
while panel (b) shows the same quantity normalized to the virial radius. The physical aperture clearly increases with halo mass and is well described by a power-law relation close to $r_{\rm cut}\propto M_{\rm halo}^{0.28}$. By contrast, the scaled aperture, $r_{\rm cut}/R_{\rm vir}$, depends only weakly on halo mass, with a best-fitting slope of about $-0.007$ per dex.

The fact that the physical aperture scales with halo mass with a slope close to one third is physically meaningful. At fixed redshift, the virial radius scales approximately as $R_{\rm vir}\propto M_{\rm halo}^{1/3}$. Therefore, if the BCG--ICL separation follows the global size of the host halo, a similar mass dependence is expected. The measured slope of the physical aperture is very close to this value, suggesting that the optimal separation radius is primarily set by the halo scale rather than by a fixed physical radius. This interpretation is further supported by the weak residual dependence on halo mass found when the
aperture is expressed in units of $R_{\rm vir}$. Once the cut is normalized by the virial radius, most of the halo-mass trend is removed. The approximate $r_{\rm cut}\propto M_{\rm halo}^{0.3}$ scaling therefore provides a natural motivation for describing the BCG--ICL separation through a halo-scaled aperture.

The shallow dependence of the physical aperture on halo mass also limits the effect of observational uncertainties in $M_{\rm halo}$. Since $R_{\rm cut}\propto M_{\rm halo}^{0.284}$, the corresponding uncertainty scales as $$\sigma_{\log R_{\rm cut}}=0.284\,\sigma_{\log M_{\rm halo}}.$$ Thus, halo-mass uncertainties of 0.2 and 0.3 dex translate into uncertainties of only 0.057 and 0.085 dex, respectively, in $\log (R_{\rm cut})$.

As a complementary parameterization, we also define $R_{\rm e,\star}$ as the projected radius enclosing half of the intrinsic BCG stellar mass, obtained from the projected Jaffe bulge combined with the exponential disk. We find $$\log_{10}(R_{\rm cut}/{\rm kpc}) = 1.522 + 0.555\,\log_{10}(R_{\rm e,\star}/{\rm kpc}).$$ The relation has an object-to-object scatter of 0.187 dex, slightly larger than the 0.168 dex scatter of the halo-mass relation measured over the same sample. Thus, $R_{\rm e,\star}$ provides a useful alternative when a reliable halo-mass estimate is unavailable, although it does not fully encode the mass and spatial distribution of the ICL and is therefore not a replacement for the primary halo-mass calibration. We also caution that the model projected half-mass radius does not need to coincide exactly with an observed half-light radius because of stellar-population gradients and observational systematics.

We then use the two best-fitting aperture prescriptions shown in Figure~\ref{fig:fig1} to quantify how well the intrinsic ICL mass is recovered. For each halo, we assign $r_{\rm cut}$ either from the physical fit or from the scaled fit, and compute the corresponding extracted ICL mass. Figure~\ref{fig:fig2} shows the ratio between this extracted value and the true ICL mass predicted by the SAM, as a function of halo mass. The blue curves show the median trends, while the dashed lines indicate the 16--84 percentile ranges; the orange horizontal line marks perfect recovery. For both aperture prescriptions, the median recovery remains close to unity, with an average scatter of $\pm 3\%$. The scatter is larger at lower halo masses and decreases toward the cluster-mass regime. This is an important result, since observational studies of the ICL are usually focused on massive groups and clusters, where our aperture prescriptions provide the most stable recovery.

Figure~\ref{fig:fig3} shows the complementary test for the BCG component. Here we measure the ratio between the BCG mass recovered inside the aperture and the true BCG mass predicted by the SAM. As expected, the aperture does not recover exactly the full BCG mass, because a small fraction of the BCG lies outside $r_{\rm cut}$ and is therefore assigned to the extracted ICL. However, this effect is small: both aperture prescriptions recover more than about $92\%$ of the BCG mass over the halo-mass range considered, even when the 16--84 percentile scatter is taken into account. Thus, the contamination of the extracted ICL by BCG stars is limited to only a few percent of the total BCG mass. The corresponding impact on the ICL fraction can nevertheless be more significant, because the ICL is much more spatially extended and on average as massive as the BCG (our SAM predicts ${\rm ICL/BCG}=1.20\pm0.02$). This explains why the ICL recovery in Figure~\ref{fig:fig2} shows a larger scatter than the BCG recovery in Figure~\ref{fig:fig3}.

For comparison, in Appendix~\ref{app:fixedapertures} we repeat the recovery test using fixed physical apertures of 30, 50, 70, and 100 kpc, similar to values commonly adopted in the literature. This test shows that fixed apertures can introduce a stronger halo-mass dependence in the recovered ICL mass, further supporting the use of a halo-scaled aperture.

It must be noted that our results might depend on the parameters used, in particular the ICL concentration and the set of profiles used. Figure~\ref{fig:fig4} shows how robust the inferred scaled aperture is with
respect to the structural assumptions adopted for the BCG--ICL system at $z=0$. In each case, we repeat the optimization procedure and measure the normalization $A_R$, namely the value of $r_{\rm cut}/R_{\rm vir}$ at the pivot halo mass $10^{14}M_\odot$. The horizontal dashed line marks the fiducial value, $A_R=0.045$ (similar to the value, 0.04, found by \citealt{marini2022}), while the shaded region shows the 16--84 percentile range obtained from bootstrap resampling of the fiducial model.

\begin{figure*}[t!]
\centering
\includegraphics[width=0.94\textwidth]{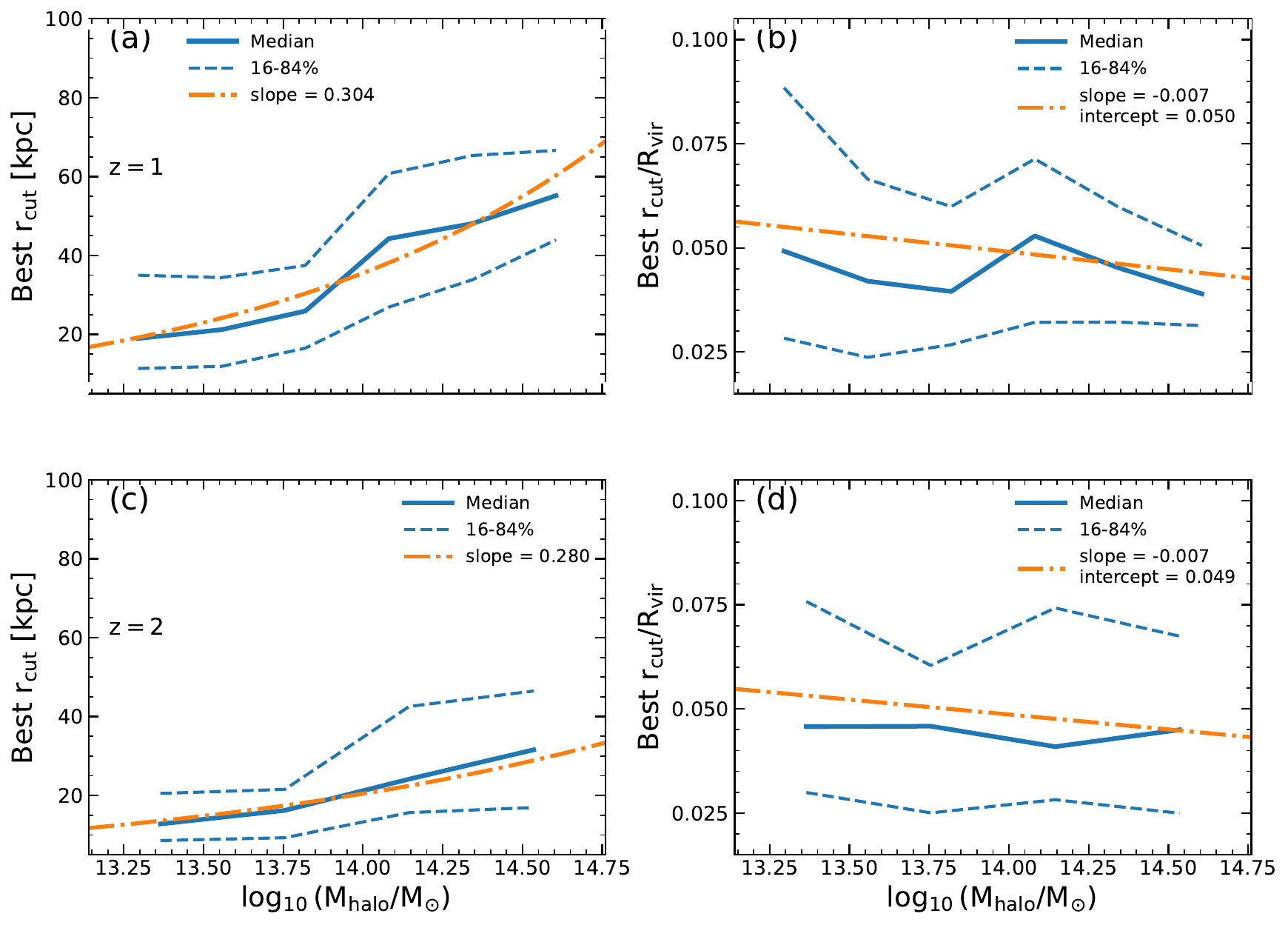}
\caption{Optimal aperture radius as a function of halo mass at $z=1$ and $z=2$. Panels (a) and (c) show the best value of $r_{\rm cut}$ in physical units for $z=1$ and $z=2$, respectively, while panels (b) and (d) show the corresponding aperture normalized to the virial radius, $r_{\rm cut}/R_{\rm vir}$. In each panel, the solid blue line shows the median relation and the dashed blue lines indicate the 16--84 percentile range. The orange dot-dashed lines show the best-fitting linear trends. The slopes reported in the legends quantify the mass dependence of the optimal aperture at each redshift. In physical units, the optimal aperture increases with halo mass, with slopes close to the expected virial scaling. In units of $R_{\rm vir}$, the dependence on halo mass is much
weaker, indicating that a halo-scaled aperture provides an approximately mass-independent description of the BCG--ICL separation at these redshifts.}
\label{fig:fig5}
\end{figure*}

\begin{figure*}[t!]
\centering
\includegraphics[width=0.94\textwidth]{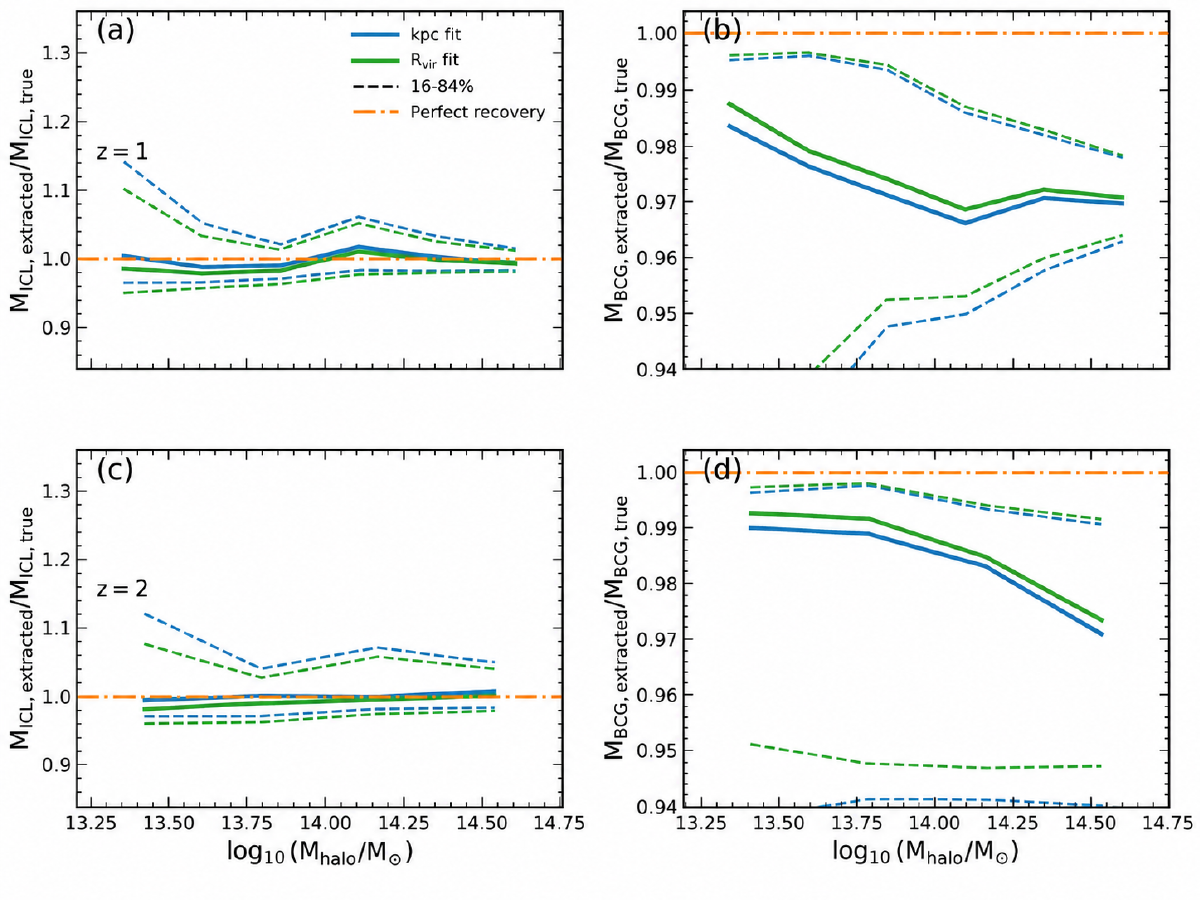}
\caption{Recovery of the ICL and BCG masses at $z=1$ and $z=2$ obtained by applying the best-fitting aperture prescriptions derived from the individual optimal cuts at each redshift. Panels (a) and (c) show the ICL recovery at $z=1$ and $z=2$, respectively, while panels (b) and (d) show the corresponding BCG recovery. For each halo, the aperture is assigned either from the fit in physical units (blue curves) or from the fit in units of the virial radius (green curves). The ICL recovery is measured as $M_{\rm ICL}^{\rm ext}/M_{\rm ICL}^{\rm true}$, where the extracted ICL includes the intrinsic ICL mass outside $r_{\rm cut}$ plus the BCG mass lying outside the same aperture. The BCG recovery is measured as $M_{\rm BCG}^{\rm ext}/M_{\rm BCG}^{\rm true}$, where the extracted BCG is the intrinsic BCG mass enclosed within $r_{\rm cut}$. Solid lines show the median relations, dashed lines indicate the 16--84 percentile ranges, and the orange dot-dashed line marks perfect recovery. Both aperture prescriptions recover the ICL and BCG masses with median values close to unity, with larger scatter in the ICL recovery than in the BCG recovery.}
\label{fig:fig6}
\end{figure*}

\begin{figure*}[t!]
\centering
\includegraphics[width=0.94\textwidth]{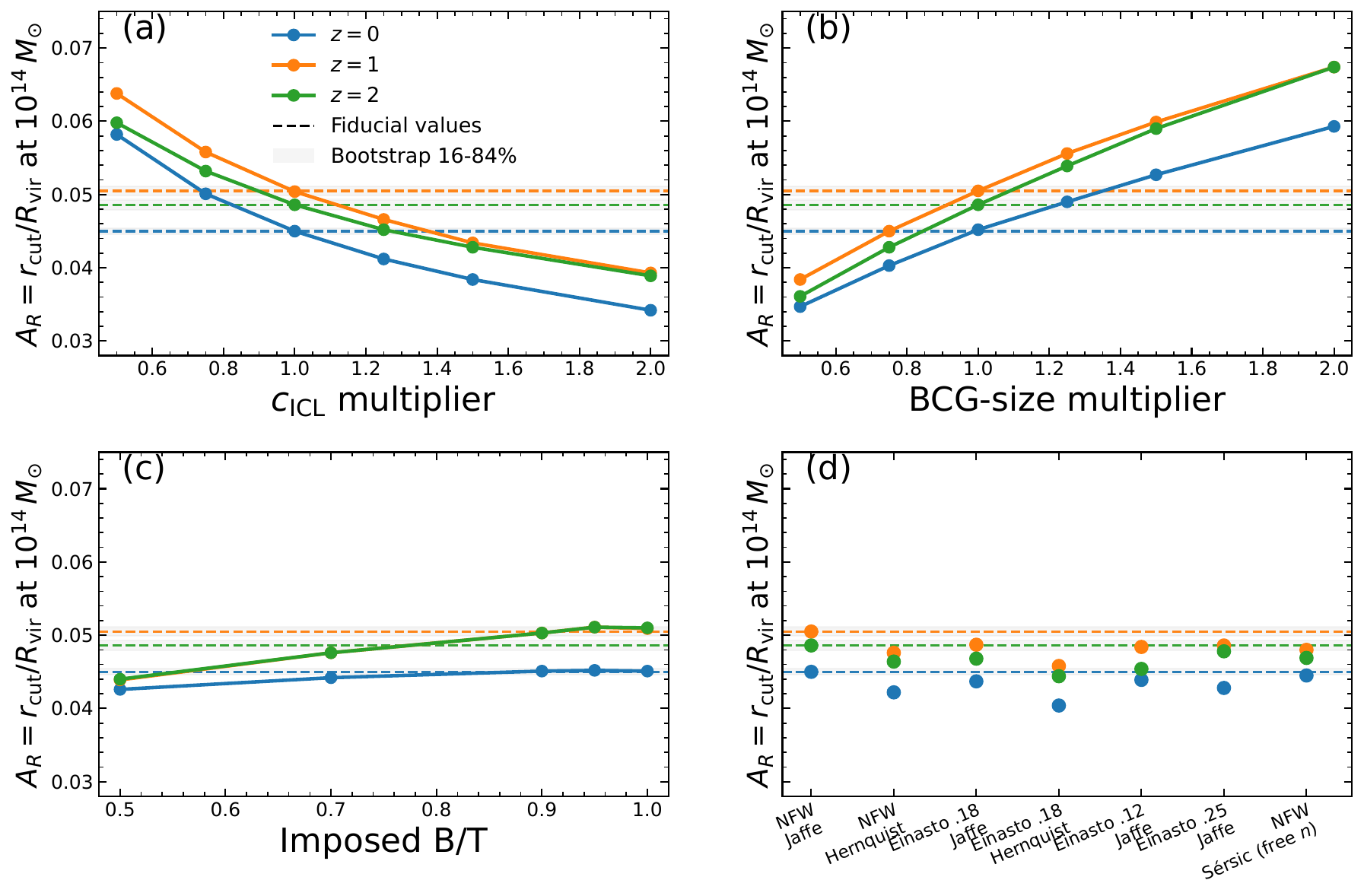}
\caption{Robustness of the scaled optimal aperture at $z=1$ and $z=2$ with respect to the assumed structural properties of the BCG and ICL components. The quantity shown is $A_R$, defined as the value of $r_{\rm cut}/R_{\rm vir}$ at the pivot halo mass $10^{14}M_{\odot}$. Orange and green curves refer to $z=1$ and $z=2$, respectively, compared with those at $z=0$ from Figure~\ref{fig:fig4} in blue. Dashed horizontal lines mark the corresponding fiducial values, while the shaded bands show the 16--84 percentile ranges obtained from bootstrap resampling of the fiducial model. Panel (a) shows the effect of varying the ICL concentration relative to the fiducial value. Increasing the concentration makes the diffuse component more centrally concentrated and therefore shifts the optimal aperture to smaller values. Panel (b) shows the effect of rescaling the characteristic size of the BCG: more extended BCG profiles require larger apertures. Panel (c) shows the dependence on the imposed bulge-to-total ratio, which changes the relative
weight of the spheroidal and disk components. Panel (d) compares different combinations of BCG and ICL profile shapes. The same qualitative trends are found at both redshifts, while the overall normalization of the optimal aperture is lower at $z=2$ than at $z=1$.}
\label{fig:fig7}
\end{figure*}

In the fiducial case, the BCG is described by a Jaffe bulge \citep{jaffe1983} plus an exponential disk \citep{freeman1970}, while the ICL follows an NFW-like profile \citep{nfw1996} \footnote{These profiles are those used by the semianalytic model \citep{contini2020b}.}. Panel (a) in Figure~\ref{fig:fig4} shows that increasing the ICL concentration moves the optimal aperture to smaller values. This is expected, because a more centrally concentrated diffuse component places a larger fraction of the ICL inside a given aperture, so the balance between missed ICL and BCG contamination is reached at smaller $r_{\rm cut}/R_{\rm vir}$. Panel (b) shows the opposite trend for the size of the BCG: when the BCG profile is made more extended, more BCG mass lies at large radii and a larger aperture is required to avoid assigning too much BCG light to the ICL. The variations of ICL concentration and BCG radial extent are explored separately here in order to isolate their individual effects on the
optimal aperture. In a physical system, however, these structural changes may be correlated. If the ICL becomes more centrally concentrated while the BCG simultaneously becomes more compact, both effects act in the same direction and shift the optimal aperture toward smaller radii. Conversely, a more extended ICL together with a more extended BCG would favour a larger aperture. The quantitative amplitude of such a coupled change depends on the relation between ICL concentration and BCG size, which is model dependent and is not imposed in the present sensitivity tests.
Panel (c) shows that increasing the imposed bulge-to-total ratio also increases the optimal aperture, although the dependence is much weaker than that obtained by changing the overall BCG size. Finally, panel (d) compares different profile families, replacing the fiducial Jaffe bulge with a Hernquist profile \citep{hernquist1990} and the fiducial NFW-like ICL with Einasto profiles \citep{einasto1965}. For the ICL component, we also consider Einasto profiles with different shape parameters, $\alpha=0.12$, $0.18$, and $0.25$. These values span different degrees of profile curvature: smaller values of $\alpha$ correspond to more gradually varying and more extended profiles, while larger values produce a stronger radial curvature. This allows us to test whether the inferred optimal aperture depends on the detailed shape of the diffuse stellar distribution.

We also test a single S\'ersic profile for the total BCG, with the S\'ersic index fitted independently for each system. The resulting optimal aperture remains close to the fiducial value, indicating that the inferred separation scale is not driven by the explicit
bulge--disk decomposition adopted in the reference model. These variations produce changes that are generally smaller than those induced by large rescalings of the ICL concentration or BCG size. Overall, the test shows that the absolute value of the optimal aperture depends on the assumed structural model, but the fiducial value remains well within the range spanned by physically motivated alternatives.

These results are broadly consistent with the picture emerging from both observational and theoretical studies, namely that the separation between the BCG and the ICL is intrinsically method dependent. Observational works based on surface-brightness limits, fixed apertures, profile decomposition, or transition radii have shown that the inferred ICL fraction can change significantly depending on the adopted definition
\citep{zibetti2005,burke2015,demaio2018,montes2018,zhang2019,kluge2021,contini2022,brough2024,mayes2026}. Our results provide a model-based explanation for this behaviour. The optimal aperture is not a universal physical radius, but instead follows the scale of the host halo, with $r_{\rm cut}\propto M_{\rm halo}^{0.3}$ and an approximately constant value in units of $R_{\rm vir}$. This naturally implies that a fixed aperture can work well only over a limited range of halo masses.

It is useful to compare the characteristic scale of our aperture with transition radii inferred observationally, although the quantities are not formally identical. Observational transition radii are generally derived from projected surface-brightness profiles, whereas our $r_{\rm cut}$ is a three-dimensional, integrated quantity. \citet{montes2021}, for example, find that the surface-brightness and colour profiles of Abell~85 become shallower beyond approximately $75\,{\rm kpc}$, which they associate with the onset of ICL dominance. Using the reported halo mass of Abell~85 in our $z=0$ relation gives $r_{\rm cut}\simeq92\,{\rm kpc}$, broadly consistent with this characteristic scale given the different definitions. Similarly, \citet{zhang2026} identify a transition at approximately $40$--$60\,{\rm kpc}$ from stacked surface-brightness profiles. This range overlaps our prediction of $r_{\rm cut}\simeq41\,{\rm kpc}$ at $M_{\rm halo}=10^{14}\,M_\odot$, increasing to approximately $79\,{\rm kpc}$ at $10^{15}\,M_\odot$.

The value found at $z=0$, $r_{\rm cut}/R_{\rm vir}\simeq 0.045$ at $10^{14}M_\odot$, is also compatible with the idea of a broad transition region rather than a sharp boundary between BCG and ICL. This is in line with studies showing that the outer BCG envelope and the diffuse light are spatially embedded and difficult to separate uniquely \citep{gonzalez2005,seigar2007,iodice2017,montes2018,kluge2021,marini2022}. In this sense,
our aperture should not be interpreted as a physical discontinuity in the stellar distribution, but as the radius where the loss of true ICL inside the aperture and the contamination from BCG stars outside the aperture compensate each other.

To quantify the radial extent over which both components contribute significantly, we define the local radial contribution of component $X$ as $q_X(r)={\rm d}M_X(<r)/{\rm d}\ln r$ and identify the 10\% overlap region through $0.1\leq q_{\rm BCG}(r)/q_{\rm ICL}(r)\leq10$. At $z=0$, the median inner and outer boundaries of this region are $12.2\,{\rm kpc}$ and $94.2\,{\rm kpc}$, with 16th--84th percentile ranges of $8.0$--$20.9\,{\rm kpc}$ and $51.4$--$153.7\,{\rm kpc}$, respectively. Its median width is therefore $\Delta r_{10}=81.7^{+52.1}_{-38.5}\,{\rm kpc}$, corresponding to $\Delta r_{10}/R_{\rm vir}=0.089$, with a 16th--84th percentile range of $0.050$--$0.141$. The BCG--ICL overlap is consequently an extended region, typically several tens of kpc wide, rather than a sharp boundary. The optimal $R_{\rm cut}$ lies within this interval for every halo in the sample, although it does not generally coincide with the local equality radius because it is defined through the integrated condition $M_{\rm BCG,out}=M_{\rm ICL,in}$.

The robustness tests further reinforce this interpretation. The inferred aperture is most sensitive to the assumed spatial extent of the BCG and to the concentration of the ICL, while changes in the analytic form of the profiles produce more moderate variations. This agrees with the general conclusion that ICL measurements depend on how the diffuse component is operationally defined, both in observations and in simulations
\citep{kluge2021,brough2024,brown2024,butler2025,brown2026}. The present implementation assumes spherical, concentric profiles centred on the host halo and therefore does not explicitly account for asymmetries, substructure, or offsets between the BCG and the halo centre. The present calibration should consequently be interpreted as most directly applicable to approximately relaxed systems, while its performance in dynamically disturbed clusters will require dedicated mock-observation or particle-based tests. The numerical calibration presented here is specific to FEGA25 and should therefore be independently tested and, if necessary, recalibrated before being applied to other semi-analytic models or hydrodynamical simulations. At the same time, the BCG recovery remains very high, with only a few percent of the BCG mass being assigned to the extracted ICL. The larger scatter in the recovered ICL fraction therefore mostly reflects the extended and low-density nature of the diffuse component, rather than a substantial loss of BCG stellar mass.

\subsection{ICL Recovery at High Redshifts}
\label{sec:redshift}

We repeat the same analysis at higher redshift, focusing on $z=1$ and $z=2$. Figure~\ref{fig:fig5} shows the equivalent of Figure~\ref{fig:fig1} at $z=1$ (panels (a) and (b)) and $z=2$ (panels (c) and (d)).
The qualitative behaviour found at $z=0$ remains valid at higher redshift. In physical units, the optimal aperture increases with halo mass, with slopes close to $M_{\rm halo}^{0.3}$ in both redshift bins. This confirms that the physical cut approximately follows the virial scaling of the host halo also at earlier epochs. When the aperture is normalized by $R_{\rm vir}$, the residual dependence on halo mass becomes very weak, comparable to that at $z=0$ (see the values of intercepts and slopes in panels (b) and (d)). Thus, most of the halo-mass dependence is again removed once the cut is expressed in units of the virial radius.

In Figure~\ref{fig:fig6}, we test how well the fitted aperture prescriptions recover the intrinsic ICL and BCG masses at $z=1$ and $z=2$, in analogy with Figures~\ref{fig:fig2} and \ref{fig:fig3} for $z=0$. Panels (a) and (c) show the ICL recovery, while panels (b) and (d) show the BCG recovery. The results are both qualitatively and quantitatively similar to those obtained at $z=0$. The BCG mass is recovered almost completely, with only a small fraction of the intrinsic BCG component assigned to the extracted ICL. The ICL recovery is also close to unity on average, although with a larger scatter, as already found in the local sample. Overall, this shows that our aperture-based method remains effective at least up to $z=2$, where both observations and theoretical predictions indicate that the ICL fraction can already be comparable to that measured at the present time \citep{joo2023,contini2024a}.

The weak mass dependence of the BCG recovery also changes slightly with redshift. At $z=0$, the recovered BCG fraction increases mildly toward larger halo masses, whereas at $z=1$ and $z=2$ it decreases toward the high-mass end. This difference reflects the evolution of the BCG structure relative to the assigned aperture rather than a reversal of the halo-mass scaling of $r_{\rm cut}$. In the high-redshift catalogues, the most massive BCGs become increasingly bulge dominated
and their characteristic bulge scale becomes larger relative to the adopted aperture. As a result, a slightly larger fraction of the BCG stellar mass lies outside $r_{\rm cut}$ and is assigned to the extracted ICL. The effect remains small, however, with the median BCG recovery remaining close to unity over the full mass range considered.

Finally, we repeat at high redshift the same robustness test performed at $z=0$. This allows us to assess whether the inferred aperture normalization remains stable against variations in the structural parameters of the BCG and ICL profiles.

Figure~\ref{fig:fig7} shows the robustness test for the high-redshift samples (compared with that at $z=0$ from Figure~\ref{fig:fig4}). As for the $z=0$ analysis, we quantify the effect of changing the structural assumptions by measuring the normalization $A_R$, i.e. the value of $r_{\rm cut}/R_{\rm vir}$ at the pivot halo mass $10^{14}M_\odot$. The orange and green curves refer to $z=1$ and $z=2$, respectively, while the horizontal dashed lines indicate the corresponding fiducial values (overplotted in blue the results at $z=0$). The shaded regions show the 16--84 percentile ranges obtained from bootstrap resampling of the fiducial model.

The qualitative behaviour is very similar to that found at $z=0$. Increasing the ICL concentration decreases the optimal aperture, because a more centrally concentrated diffuse component places more ICL mass inside a given radius. In contrast, increasing the characteristic size of the BCG shifts the optimal cut to larger values, since more BCG mass is distributed at large radii and a larger aperture is required to reduce the contamination of the extracted ICL. The dependence on the imposed bulge-to-total ratio is weaker, but it follows the same general trend: more centrally dominated BCG models require slightly larger values of $A_R$. Finally, variations in the analytic form of the profiles produce moderate changes compared to the effect of large rescalings of the ICL concentration or BCG size.

The most important result is that the normalization of the optimal scaled aperture remains comparable to that at $z=0$. This is particularly relevant because both observations and models suggest that the ICL can already represent a significant fraction of the stellar light at $z>1$ \citep{joo2023,werner2023,contini2024a}, while its separation from the BCG is even more challenging than in the local Universe. In this context, the stability of the trends in Figure~\ref{fig:fig7} indicates that an aperture tied to the halo scale provides a useful and physically motivated way to describe the BCG--ICL transition also at high redshift.

The weak and mildly non-monotonic evolution of the aperture normalization reflects competing changes in the BCG and ICL structures. At the pivot halo mass, $A_R$ increases from 0.045 at $z=0$ to 0.050 at $z=1$, and remains nearly unchanged at $z=2$, where $A_R=0.049$. From $z=0$ to $z=1$, the median ICL concentration decreases from $c_{\rm ICL}\simeq9.98$ to 8.03, while $M_{\rm ICL}/M_{\rm BCG}$ decreases from about 1.25 to 1.03. Both effects shift the balance condition $M_{\rm BCG,out}=M_{\rm ICL,in}$ toward larger $r_{\rm cut}/R_{\rm vir}$. Indeed, replacing the $z=1$ ICL concentration or ICL-to-BCG mass ratio with their $z=0$ trends lowers the inferred normalization to 0.047 and 0.046, respectively. At $z=2$, the ICL becomes even less concentrated, $c_{\rm ICL}\simeq6.56$, but this is compensated by a more compact and less bulge-dominated BCG: $a/R_{\rm vir}$ decreases from $1.75\times10^{-3}$ to $1.38\times10^{-3}$, while $B/T$ decreases from 0.84 to 0.64. These structural changes reduce the amount of BCG mass at large radii and prevent a further increase in the optimal aperture. The near constancy of $A_R$ between $z=1$ and $z=2$ therefore results from a compensation between the decreasing ICL concentration and the evolving BCG structure, rather than from an absence of structural evolution.

\section{Conclusions}
\label{sec:conclusion}

In this work we have introduced an aperture-based method to separate the brightest cluster galaxy (BCG) from the intracluster light (ICL) using the intrinsic components predicted by our semi-analytic model. The BCG is modelled as a central stellar component, while the ICL is treated as a diffuse component associated with the host halo. For each system, we define an optimal aperture, $r_{\rm cut}$, by requiring the extracted ICL mass to be as close as possible to the intrinsic ICL mass. In practice, this corresponds to the radius at which the amount of BCG mass lying outside the aperture balances the amount of ICL mass enclosed within it. This provides a physically motivated way to connect
an observationally inspired aperture definition to the intrinsic BCG--ICL decomposition of the model.

Our main results can be summarized as follows.

First, at $z=0$, the optimal aperture in physical units increases with halo mass with a slope close to $M_{\rm halo}^{1/3}$. This scaling is naturally expected if the BCG--ICL transition follows the global size of the host halo. When the aperture is expressed in units of the virial radius, $r_{\rm cut}/R_{\rm vir}$, most of the halo-mass dependence is removed. This makes the scaled aperture a more stable description of the BCG--ICL separation than a fixed physical radius.

Second, the aperture prescriptions derived from the individual optimal cuts recover both components well. The BCG mass is recovered almost completely, with only a small fraction ($4\%$ on average) of the intrinsic BCG assigned to the extracted ICL. The ICL recovery shows a larger scatter, especially at lower halo masses, but its median value remains close to unity. This indicates that both the physical and halo-scaled prescriptions provide an effective recovery of the intrinsic BCG and ICL masses, with the scaled aperture having the advantage of a weaker dependence on halo mass.

Third, the same qualitative behaviour is found at higher redshift. At $z=1$ and $z=2$, the physical aperture still scales approximately as $M_{\rm halo}^{0.3}$, while the scaled aperture remains nearly independent of halo mass. Moreover, we find no significant evolution of both intercept and slope in the case of the normalized aperture as a function of redshift, which is important considering the growing amount of ICL studies at high redshifts.

Finally, we have tested the robustness of the inferred aperture against changes in the structural assumptions of the model. The absolute value of the optimal cut is most sensitive to the ICL concentration and to the characteristic size of the BCG, while variations in the analytic form of the adopted profiles produce more moderate changes. Importantly, the main trends with halo mass and redshift are preserved, indicating that our conclusions are not driven by a particular choice of fiducial parameters.

Overall, our results show that an aperture tied to the virial radius provides a simple and physically motivated way to describe the BCG--ICL separation. While no single aperture can represent a unique physical boundary between the two embedded components, the method proposed here identifies the radius that minimizes the bias in the recovered ICL mass. This makes it a useful tool for interpreting observational aperture-based measurements and for connecting them to the intrinsic BCG and ICL components predicted by galaxy formation models.


\section*{Acknowledgements}
The authors acknowledge support from the Korean National Research Foundation (RS-2022-NR070872 and RS-2025-00514475).

\section*{Data Availability}
The catalogs used to generate the figures and support the scientific analysis presented in this work are publicly available on Zenodo under a Creative Commons Attribution 4.0 International (CC BY 4.0) license: \dataset[doi:10.5281/zenodo.24295270] {https://doi.org/10.5281/zenodo.24295270}. The repository also includes a README file providing information on the contents of the catalogs and
instructions for their use.

\appendix

\section{Fixed physical apertures}
\label{app:fixedapertures}

Several observational and numerical studies adopt fixed physical apertures to separate the BCG from the diffuse stellar component. To compare this approach with the optimal apertures discussed in the main text, we repeat the recovery test using four commonly adopted fixed cuts: 30, 50, 70, and 100 kpc. The results are shown in Figure~\ref{fig:appfixedapertures}.

Fixed physical apertures do not provide a uniform recovery of the intrinsic components over the full halo-mass range. For a given aperture, the recovered ICL mass reflects the balance between two competing effects: contamination from BCG stars lying outside the aperture and the loss of true ICL enclosed within it. Small apertures leave a larger fraction of the outer BCG outside the cut and can therefore lead to an overestimate of the ICL mass, particularly at the high-mass end. Increasing the aperture reduces this contamination, but simultaneously includes a larger fraction of the true ICL inside the BCG aperture, which can produce an underestimate of the recovered ICL mass, especially in lower-mass halos. Consequently, increasing the aperture does not monotonically improve the recovery, but instead shifts the halo-mass scale at which the bias changes from overestimation to underestimation.

The $70\,{\rm kpc}$ aperture provides a useful observational comparison. \citet{gonzalez2021} adopt a transition at approximately this radius for MACS~J1149.5+2223. In our experiment, a fixed $70\,{\rm kpc}$ cut recovers nearly all of the intrinsic ICL mass in net at the high-mass end, while the median recovered fraction is approximately $0.94$ at $M_{\rm halo}\sim10^{14}\,M_\odot$ and decreases to about $0.84$ at the lowest halo masses considered. Thus, depending on halo mass, the same fixed aperture can produce a negligible bias in massive clusters but underestimate the recovered ICL by roughly $15\%$ in lower-mass systems. We stress that this is the net recovered ICL fraction, $M_{\rm ICL}^{\rm ext}/M_{\rm ICL}^{\rm true}$, and not simply the fraction of intrinsic ICL geometrically lying outside the aperture, since BCG stars outside the same radius also contribute to the extracted diffuse component.

The recovered BCG mass is less sensitive to the aperture choice and remains relatively close to the intrinsic value over most of the halo-mass range. The main systematic uncertainty introduced by fixed apertures therefore affects the inferred ICL mass. This comparison further motivates the use of an aperture tied to the halo size, which substantially reduces the halo-mass dependence seen for fixed physical cuts.

\label{app:fixedapertures}
\setcounter{figure}{0}
\renewcommand{\thefigure}{A\arabic{figure}}

\begin{figure}
\centering
\includegraphics[width=\columnwidth]{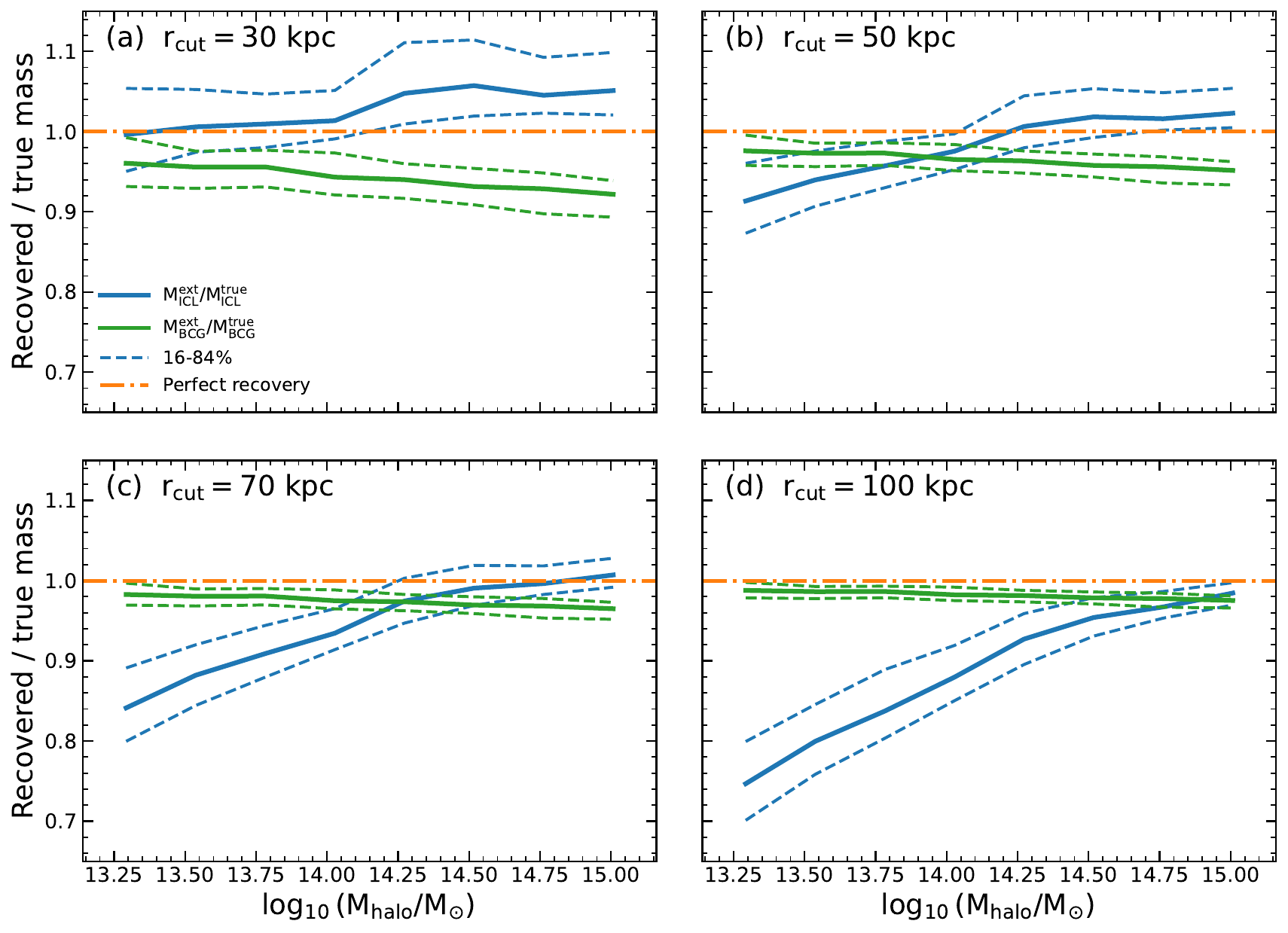}
\caption{
Recovery of the ICL and BCG masses at $z=0$ using fixed physical apertures of 30, 50, 70, and 100 kpc. Each panel corresponds to a different value of $r_{\rm cut}$. Blue curves show $M_{\rm ICL}^{\rm ext}/M_{\rm ICL}^{\rm true}$, green curves show $M_{\rm BCG}^{\rm ext}/M_{\rm BCG}^{\rm true}$, and dashed curves indicate the 16--84 percentile ranges. The orange dot-dashed line marks perfect recovery.
}
\label{fig:appfixedapertures}
\end{figure}

\bibliography{paper}{}
\bibliographystyle{aasjournalv7}



\end{document}